\documentstyle[12pt]{article}
\input epsf

\def\be{\begin{equation}}
\def\ee{\end{equation}}
\def\balpha{ {\mbox{\boldmath $\alpha$}}}
\def\bbeta{ {\mbox{\boldmath $\beta$}}}
\def\bgamma{ {\mbox{\boldmath $\gamma$}}}
\def\bsigma{ {\mbox{\boldmath $\sigma$}}}

\def\fun#1#2{\lower3.6pt\vbox{\baselineskip0pt\lineskip.9pt
  \ialign{$\mathsurround=0pt#1\hfil##\hfil$\crcr#2\crcr\sim\crcr}}}

\begin{document}

\thispagestyle{empty}
\rightline{CU-TP-785}
\rightline{FERMILAB-CONF-96/349-T}
\rightline{hep-th/9610065}
\vskip 2cm

\centerline{\Large\bf Massless and Massive Monopoles Carrying}
\centerline{{\Large\bf Nonabelian Magnetic Charges}\footnote{Talk
delivered at the International Seminar ``Quarks-96'', Yaroslavl,
Russia,  May 1996}  }
\vskip 1.0cm
\centerline{Erick J. Weinberg} \vskip 2mm
\centerline{\it Physics Department, Columbia University, New York, NY
10027, USA\footnote{Permanent address} and }
%\centerline{ and}
\centerline{\it Fermi National Accelerator Laboratory, P.O. Box 500,
Batavia, IL 60510, USA}
\vskip 1.0cm
\centerline{\bf ABSTRACT}
\vskip 5mm
\begin{quote}
{ The properties of BPS monopoles carrying nonabelian magnetic charges
are investigated by following the behavior of the moduli space of
solutions as the Higgs field is varied from a value giving a purely
abelian symmetry breaking to one that leaves a nonabelian subgroup of
the gauge symmetry unbroken.  As the limit of nonabelian unbroken
symmetry is reached, some of the fundamental abelian monopoles remain
massive but acquire nonabelian magnetic charges.  The BPS mass formula
indicates that others should because massless in this limit.  These do
not correspond to distinct solitons, but instead are manifested as a
``nonabelian cloud'' surrounding the massive monopoles, with their
position and phase degrees of freedom being transformed into
parameters characterizing the cloud.}

\end{quote}

\section{Introduction}

Magnetic monopoles have long been the object of great interest.  
This is due in part to their role as predicted, but as yet undiscovered,
particles in all grand unified theories.  Beyond this, however, the
monopoles of spontaneously broken gauge theories are of interest as
examples of particles, arising from classical soliton solutions, that
are in a sense complementary to the elementary quanta
of the theory.  It is particularly striking that this soliton-quanta
complementarity mirrors the magnetic-electric duality of Maxwell's
equations.  This connection is made more concrete in the conjecture by
Montonen and Olive \cite{dual} that certain theories may possess an
exact magnetic-electric duality that exchanges solitons with elementary
quanta and weak with strong coupling.

In this talk I will describe some research, done in collaboration with
Kimyeong Lee and Piljin Yi \cite{ourpaper}, concerning monopoles that
carry nonabelian magnetic charges; i.e., those whose long-range
magnetic field transforms nontrivially under an unbroken nonabelian
subgroup of the gauge symmetry of the theory.  Just as the elementary
quanta carrying nonabelian electric-type charges give rise to
phenomena that are not seen with purely abelian charges, one might
expect nonabelian magnetic charges to display interesting new
features.  Indeed, past investigations have discovered some curious
properties associated with the long-range fields of such monopoles.
For example, attempts to obtain chromodyons \cite{abouel}, objects
with both electric and magnetic nonabelian charges, by applying
time-dependent global gauge rotations are frustrated by topological
obstructions \cite{manohar} to the definition of such rotations in the
presence of a nonabelian magnetic charge.  Also, it has been shown
\cite{brandt} that the large-distance behavior of their Coulomb
magnetic field can lead to instabilities in monopoles with more than a
minimal nonabelian magnetic charge.  New issues concerning these
objects are raised by the duality conjecture.  One, with which I will
be particularly concerned in this talk, is the nature of the objects
that are the magnetic counterparts of the massless nonabelian gauge
bosons.  Duality suggests that these should be massless, but it is not
at all clear how one would obtain a zero energy soliton.

As I will describe below, our strategy was to start with a theory
whose gauge symmetry is spontaneously broken to a purely abelian
subgroup, and then to follow the behavior of the classical monopole
solutions as the asymptotic Higgs field is varied to one of the
special values that correspond to a nonabelian unbroken symmetry. To
avoid the pathologies associated with the long-range behavior of
nonabelian magnetic fields, we focused on systems, generally
containing more than one monopole, whose total magnetic charge was
purely abelian \cite{nelson}.  Throughout we worked in the
Bogomolny-Prasad-Sommerfield (BPS) limit \cite{bps}, with an adjoint
representation Higgs field $\Phi$.  In addition, we made extensive use
of the moduli space approximation \cite{geodesic}, in which the low
energy dynamics of interacting monopoles is reduced to that of a small
number of collective coordinates.

The remainder of this talk is organized as follows.  In Sec.~2, I
review some of the properties of monopole and multimonopole solutions
with both $SU(2)$ and larger gauge groups.  The moduli space
approximation is described in Sec.~3.  In Sec.~4, I illustrate the 
behavior of monopoles as one goes from an abelian to a nonabelian
unbroken symmetry, using an $SO(5)$ example for which it is possible
to carry out explicit calculations.  The extension to other gauge
groups is discussed in Sec.~6.  Section 7 contains come concluding
remarks.

\section{Multimonopole solutions with SU(2) and larger gauge groups}

For an $SU(2)$ gauge theory whose symmetry is broken to $U(1)$ by a
triplet Higgs field $\Phi$, the asymptotic magnetic field is 
\be  
    B_i^a =  {g{\hat r}_i \over 4\pi r^2} {\Phi^a \over |\Phi|} 
\ee
with $g$ quantized in integer multiples of $4\pi/e$.  The BPS solution
carrying a single unit of magnetic charge is spherically symmetric and
can be written in the form \cite{bps}
\begin{eqnarray}
   \Phi^a &=&  {\hat r}^a H(r)   \nonumber \\
   A_i^a  &=&  \epsilon_{aim}{\hat r}^m A(r)\, ,
\end{eqnarray}
where $v$ is the asymptotic magnitude of the Higgs field and 
\begin{eqnarray}
   A(r) &=& {v \over \sinh evr} - {1\over er} \nonumber \\
   H(r) &=& v \coth evr  - {1\over er}\, .
\label{AHdef}   
\end{eqnarray}

    The BPS solutions with $n>1$ units of magnetic charge are all
naturally interpreted as multimonopole solutions.  Their energy is
precisely $n$ times the mass of the unit monopole, indicating that the
attractive long-range force mediated by the Higgs field (which is
massless in the BPS limit) exactly cancels the Coulomb magnetic
repulsion between static monopoles.  Index theory methods show that
after gauge fixing there are $4n$ linearly independent normalizable
zero modes about any such solution \cite{ejw}.  Hence, the moduli space of
solutions is parameterized by $4n$ variables, which can be taken to be
three position coordinates and one $U(1)$ phase for each of the
component monopoles; allowing these parameters to vary with time
endows the individual monopoles with linear momentum or electric
charge.

    Now consider the case of an arbitrary gauge group $G$ of rank $r$.
Its generators can be chosen to be $r$ commuting operators $H_i$ that
span the Cartan subalgebra, together with raising and lowering
operators $E_{\mbox{\boldmath $\alpha$}}$ associated with the roots
$\mbox{\boldmath $\alpha$}$.  The nature of the symmetry breaking is
determined by the Higgs field.  Its asymptotic value $\Phi_0$ in some
fixed direction can be chosen to lie in the Cartan subalgebra, thus
defining a vector $\bf h$ by
\begin{equation}
           \Phi_0 = {\bf h \cdot H}\, .
\end{equation}
Maximal symmetry breaking (MSB), to the subgroup $U(1)^r$, occurs
if $\bf h$ has nonzero inner products with all of the root vectors.
If instead there are $k\ge 1$ root vectors orthogonal to $\bf h$, then
the sublattice formed by these is the root lattice for a rank $k$
semisimple group $K$ and there is a nonabelian unbroken symmetry (NUS)
$K\times U(1)^{r-k}$.

    Now recall that one can choose as a basis for the root lattice a
set of $r$ simple roots with the property that all other roots are
linear combinations of these with integer coefficients all of
the same sign.  It turns out to be particularly convenient to choose 
the simple roots so that their inner products with $\bf h$ are all
nonnegative.  In the MSB case this uniquely determines the simple
roots, which we denote by $\mbox{\boldmath $\beta$}_a$.  In the NUS
case we denote by $\mbox{\boldmath $\gamma$}_j$ the simple roots
orthogonal to $\bf h$ (i.e., the simple roots of $K$) and write the
remainder as $\mbox{\boldmath $\beta$}_a$.  In this latter case the
choice of simple roots is not unique; the various possibilities are
related by the Weyl group of $K$.

     The quantization conditions on the magnetic charge can now be
easily written down.  The asymptotic magnetic field must commute with
$\Phi$; hence, we can require that in the direction chosen to define
$\Phi_0$ it be of the form
\begin{equation}
    B_i = {{\hat r}_i \over 4\pi r^2}\, {\bf g\cdot H}\, .
\end{equation}
Topological arguments \cite{topology} then imply that
\begin{equation}
    {\bf g} = {4\pi \over e} \left[\sum_{a} n_a 
{\mbox{\boldmath $\beta$}}_a^* 
         + \sum_{j} q_j {\mbox{\boldmath $\gamma$}}_j^* \right],
\label{gcoeff}
\end{equation} 
where  
\begin{equation}
    {\mbox{\boldmath $\alpha$}}^* = {{\mbox{\boldmath $\alpha$}} \over 
{\mbox{\boldmath $\alpha$}}^2}
\end{equation} 
is the dual of the root ${\mbox{\boldmath $\alpha$}}$.
The integers $n_a$ are the topologically conserved charges.  For a given
solution they are uniquely determined and gauge-invariant, even though 
the corresponding ${\mbox{\boldmath $\beta$}}_a$ may not be.
The $q_j$ are also integers, but are neither gauge-invariant nor
conserved. 

      Consider first the MSB case.  The energy
of any BPS solution is 
\be
    M = {\bf g\cdot h} = \sum_a n_a \left( {4\pi \over e} {\bf h}
        \cdot{\mbox{\boldmath $\beta$}}_a \right) 
\label{mass}
\ee
while the number of normalizable zero modes (after gauge fixing) is
\cite{erick} 
\begin{equation}
    p = 4 \sum_a n_a \, .
\label{MSBmodes}
\end{equation} 
These results suggest that, in analogy with the $SU(2)$ case, all
solutions might be viewed as composed of a number of fundamental
monopoles, each with a single unit of topological charge, 
that have no internal degrees of freedom.
In fact,
these fundamental monopole solutions are easily constructed.  Any root
$\balpha$ defines an $SU(2)$ subgroup with generators
\begin{eqnarray}
t^1({\mbox{\boldmath $\alpha$}}) &=& \frac{1 }{ \sqrt{2
{\mbox{\boldmath $\alpha$}}^2}} (E_{\mbox{\boldmath $\alpha$}} +
E_{-{\mbox{\boldmath $\alpha$}}})                  \nonumber \\
t^2({\mbox{\boldmath $\alpha$}}) &=& -\frac{i }{ \sqrt{2{
\mbox{\boldmath $\alpha$}}^2}} (E_{\mbox{\boldmath $\alpha$}} -
E_{-{\mbox{\boldmath $\alpha$}}})                 \nonumber \\
t^3({\mbox{\boldmath $\alpha$}}) &=&  {\mbox{\boldmath $\alpha$}}^* 
\cdot  {\bf H} \, .
\label{tripletdef}
\end{eqnarray}
If $A^s_i({\bf r}; v)$ and $\Phi^s({\bf r}; v)$ give the $SU(2)$
solution corresponding to a Higgs expectation value $v$, then 
\begin{eqnarray}
   A_i({\bf r})  &=&  \sum_{s=1}^3 A_i^s({\bf r}; {\bf h}\cdot 
{\mbox{\boldmath $\beta$}}_a) t^s({\mbox{\boldmath $\beta$}}_a) \nonumber \\
   \Phi({\bf r})  &=&  \sum_{s=1}^3 \Phi^s({\bf r}; {\bf h}\cdot 
{\mbox{\boldmath $\beta$}}_a) t^s({\mbox{\boldmath $\beta$}}_a)   
   + ({\bf h} - {\bf h}\cdot {\mbox{\boldmath $\beta$}}_a^* \,
{\mbox{\boldmath $\beta$}})\cdot {\bf H} 
\label{embedding}
\end{eqnarray}
is the fundamental monopole corresponding to the root
${\mbox{\boldmath $\beta$}}_a$. It carries topological charge 
\begin{equation}
     n_b = \delta_{ab}   
\end{equation}
and has a mass
\begin{equation}
   m_a ={4\pi \over e} {\bf h}\cdot {\mbox{\boldmath $\beta$}}_a^*\,  .
\end{equation}
Finally, there are only four zero modes about this solution, three
corresponding to spatial translations and one to global rotations by
the $U(1)$ generated by $\bbeta_a \cdot {\bf H}$.  (The other $r-1$
unbroken $U(1)$ factors leave the solution invariant.)

\vskip 5mm
\begin{center}
\leavevmode
\epsfysize =2in\epsfbox{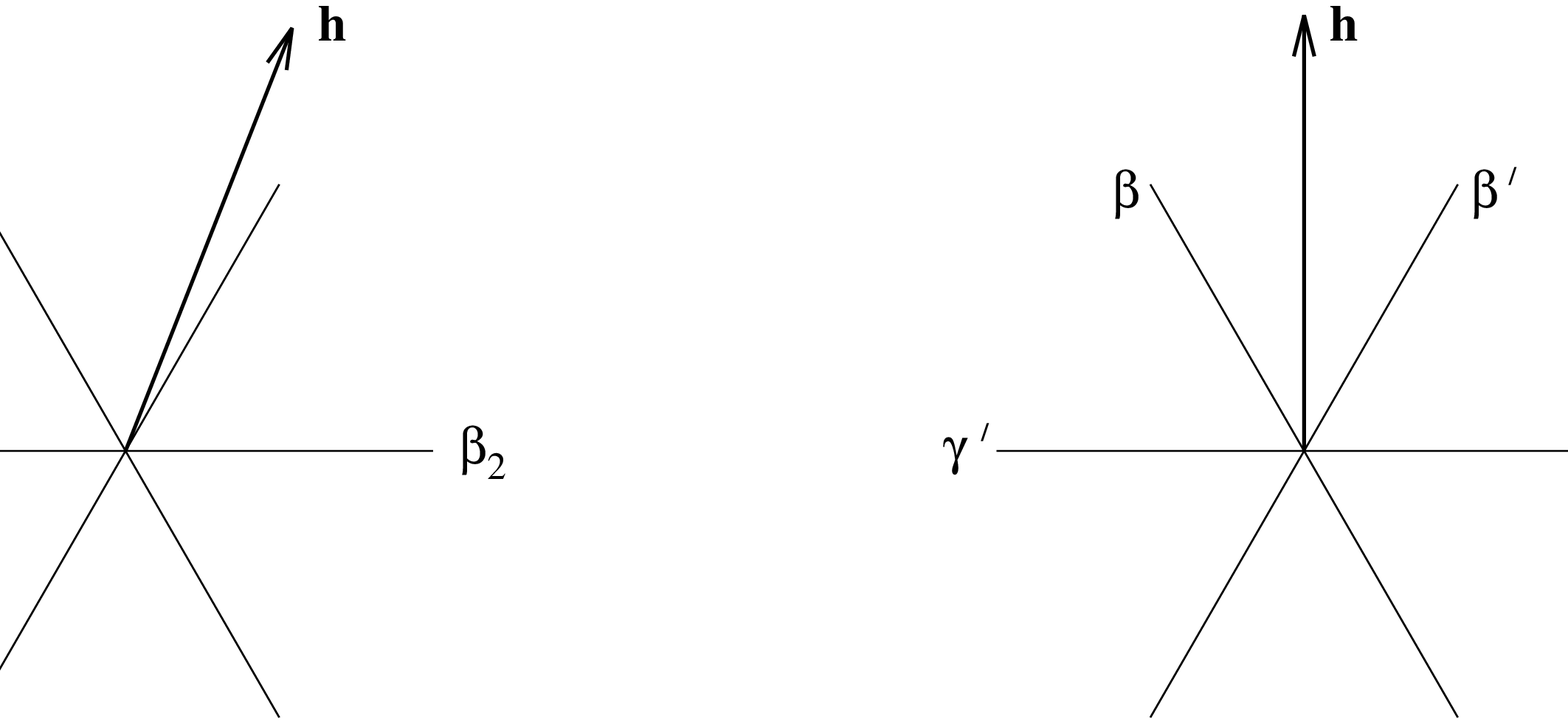}
\end{center}
\vskip 8mm
\begin{quote}
{\bf Figure 1:} {\small
The root diagram of $SU(3)$. With the Higgs vector $\bf h$ oriented as
in (a) the gauge symmetry is broken to $U(1)\times U(1)$, while with
the orientation in (b) the breaking is to $SU(2) \times U(1)$.}
\end{quote}
\vskip 0.38cm

    As a concrete example, consider the case of $SU(3)$ broken to
$U(1)\times U(1)$.  With $\bf h$ as indicated in the root diagram of
Fig.~1a, the the prescription described above gives the two simple
roots labelled $\bbeta_1$ and $\bbeta_2$.  Each of these defines an
embedding of the $SU(2)$ unit monopole about which there are four
normalizable zero modes.  One could also use the root
$\bbeta_1+\bbeta_2$ to define an embedding of the $SU(2)$ monopole.
At first glance, one might interpret the resulting spherically
symmetric solution as a third type of fundamental soliton.  However,
there are eight, rather than four, zero modes about this solution,
indicating that it is just a special case of an eight-parameter family
of two-monopole solutions.  Note, for later reference, that this
symmetric superposition of two fundamental monopoles has a
core radius that is smaller than that of either of its components.
This can be understood by noting that the presence of one monopole
changes the magnitude of the vector boson mass that determines the
core radius of the other.

     Now let us turn to the NUS case, where an unbroken nonabelian
symmetry survives.  The energy of a solution, given again by
Eq.~(\ref{mass}), depends only on the $n_a$, not on the $q_i$.  For
technical reasons associated with the presence of long-range
nonabelian fields, the index theory methods used to count zero modes
can only be used if the magnetic charge is purely abelian (i.e., if
${\bf g}\cdot \bgamma_i =0$ for all $\bgamma_i$); when these methods
can be applied, Eq.~(\ref{MSBmodes}) for the number of zero modes is
replaced by
\cite{erick2} 
\begin{equation}
    p = 4 \left[ \sum_a n_a + \sum_j q_j \right] .
\label{NUSmodes}
\end{equation}

    Our experience with the MSB case might lead us to expect that the
NUS solutions would also have a simple interpretation in terms of
fundamental objects without internal degrees of freedom.  However,
there are several difficulties with this.  First, while
Eq.~(\ref{NUSmodes}) suggests that there should be a fundamental
monopole for each of the simple roots, the mass formula implies that
those corresponding to the $\bgamma_i$ should be massless, which is
impossible for a classical soliton of this theory.  Second, since 
some of the monopole solutions obtained from the $\bbeta_a$ transform
nontrivially under the unbroken nonabelian factor $K$, one might have
expected the corresponding $n_a$ to appear with a coefficient greater
than 4 in Eq.~(\ref{NUSmodes}).  Finally, because the $q_j$ are not
invariant, the number of component monopoles in a solution would
depend on the choice of simple roots; a change of basis could even
turn an apparently fundamental solution into a composite one.

    The NUS version of the $SU(3)$ example referred to previously is
shown in Fig.~1b.  With $\bf h$ as indicated, the unbroken subgroup is
$SU(2)\times U(1)$.  The simple roots can be chosen to be either the
pair labelled $\bbeta$ and $\bgamma$ or the pair labelled $\bbeta'$
and $\bgamma'$.  As in the MSB case, a solution can be obtained by
embedding the $SU(2)$ monopole using the subgroup defined by $\bbeta$;
a gauge-equivalent solution is obtained by using $\bbeta'$.  Although
the index methods fail, the normalizable zero modes about these two
solutions can be found by direct solution of the differential
equations.  One finds that there are still only four such modes; the
additional $SU(2)$ modes that might have been expected are
nonnormalizable.  Finally, as indicated above, there is no monopole
solution corresponding to $\bgamma$; making the substitution $\bbeta_a
\rightarrow \bgamma$ in Eq.~(\ref{embedding}) gives simply the vacuum
solution.

\section{The moduli space approximation}

The BPS multimonopole and multidyon solutions describe configurations
whose component objects are all at rest relative to one another and
all have the same ratio of electric to magnetic charges.  One might
expect that the solutions for objects with small relative velocities
or arbitrary small electric charges, although not truly BPS, would in
some sense be approximately BPS.  This idea is formulated more
precisely in the moduli space approximation \cite{geodesic}.  Let us
work in $A_0=0$ gauge and adopt a notation where $\Phi=A_4$ and
$\partial_4=0$.  Now let $A_a^{BPS}({\bf r}, z)$ ($a=1,\cdots ,4$)
denote a family of gauge-inequivalent static BPS solutions
parameterized by $n$ collective coordinates $z_j$.  The moduli space
approximation consists of assuming that for any $t$ the fields can be
approximated by a configuration $A_a({\bf r},t)$ that is
gauge-equivalent to one of these BPS solutions; i.e.,
\begin{equation}
    A_a({\bf r},t) = U^{-1}({\bf r},t)\, A_a^{BPS}({\bf r}, z(t))  \,
     U({\bf r},t)  -i U^{-1}({\bf r},t) \,\partial_a U({\bf r},t) \, . 
\label{modspaceansatz}
\end{equation}
(There is a subtle point here.  Because we are working in $A_0=0$
gauge, only time-independent gauge transformations are at our
disposal; this is why there are no time derivatives of $U$ in
Eq.~(\ref{modspaceansatz}).  However, since
the gauge transformation relating $A_a$ to $A_a^{BPS}$ might be
different at one time than at another, we have to allow $U$ to be
$t$-dependent.   
Thus, $U({\bf r},t)$ should be understood as a family of static gauge
transformations parameterized by $t$.)
Given this Ansatz, the time derivatives of the
fields must be of the form
\be
   \dot A_a = \dot z_j {\partial A_a \over \partial z_j} + D_a
       (\dot z_j \epsilon_j ) \equiv \dot z_j\, \delta_j A_a
\label{Adot}
\ee
where the gauge transformation generated by $\dot z_j
\,\epsilon_j({\bf r})$ arises from the time derivative of $U$.  This
gauge function is determined by Gauss's law,
\be
   0  =  D_a \dot A_a = \dot z_j\, D_a (\delta_j A_a ) \, .
\ee
This equation that shows the $\delta_j A_a$ are simply the
background gauge zero modes about the BPS solution \cite{jerome}.

The $A_0=0$ gauge Lagrangian is
\be 
    L = \int d^3 r \, {\rm tr} \left[{1\over 2} \dot A_i^2 +{
1\over 2} \dot\Phi^2
  + {1\over 4} F_{ij}^2 + {1\over 2} D_i\Phi^2 \right] \, .
\ee 
With the Ansatz (\ref{modspaceansatz}), 
the contribution of the last two terms
is simply the BPS energy and therefore independent of time.  Using
Eq.~(\ref{Adot}) to rewrite the first two terms, we obtain
\begin{equation} 
     L = {1\over 2} g_{ij}(z)\,\dot z_i \dot z_j  + {\rm constant}
\label{modspacelag}
\end{equation} 
where 
\begin{equation}
    g_{ij}(z) = \int d^3x \,{\rm tr}\, 
         \left(\delta_i A_a \,\,\delta_j A_a \right) .
\label{modetometric}
\end{equation}
Thus, the dynamics of the fields has been reduced to that of a point
particle in geodesic motion on an $n$-dimensional moduli space with metric
$g_{ij}$(z). 

     If the full family of solutions for a given magnetic charge is
known, then it is a straightforward (at least in principle) matter to
obtain a complete set of background gauge zero modes and then
substitute into Eq.~(\ref{modetometric}) to obtain the moduli space
metric.  In most cases, however, an explicit solution is not
available.  Nevertheless, it may still be possible to determine
$g_{ij}$.  For example, the isometries of the moduli space that are
implied by the space-time and internal symmetries of the theory,
together with the requirement that the space be hyperk\"ahler (this can
be shown to follow from the properties of the BPS equations
\cite{atiyah}) may be 
sufficient to uniquely determine the metric; this was the approach
used by Atiyah and Hitchin to determine the two-monopole moduli space
metric for the $SU(2)$ theory \cite{atiyah}.  Another approach works
for arbitrary 
magnetic charge, but only in the region of moduli space corresponding
to widely separated monopoles.  Here one uses the fact that the
long-range interactions between monopoles are well understood, and
determines the metric by the requirement that the Lagrangian
(\ref{modspacelag}) reproduce these interactions \cite{gary}.

     In particular, the moduli space and its metric are known for all
cases with only two fundamental monopoles when there is maximal
symmetry breaking.  If the fundamental monopoles are of the same type,
then the solutions are essentially embeddings of $SU(2)$ solutions and
the moduli space is the Atiyah-Hitchin manifold.  If the fundamental
monopoles are distinct and correspond to orthogonal simple roots, then
there are no interactions between the monopoles and the moduli space
is simply a direct product of one-monopole moduli spaces.  The final
possibility is that the fundamental monopoles correspond to distinct
simple roots $\bbeta$ and $\bgamma$ with a nonzero inner product
$\lambda = -2\bbeta^* \cdot \bgamma^* >0$.  As described by Kimyeong
Lee in his talk at this conference
\cite{kmltalk}, the 
moduli space for this case can be determined by a combination of the
two methods described above.  It is of the form \cite{klee}
\begin{equation}
   {\cal M} = R^3 \times {R^1 \times {\cal M}_0 \over Z}
\label{space}
\end{equation} 
where the $R^3$ factor corresponds to the center-of-mass position and the
$R^1$ to an overall $U(1)$ phase angle $\chi$, while
${\cal M}_0$, corresponding to the relative coordinates, 
is the Taub-NUT space with metric 
\begin{equation}
   ds_0^2 = \left(\mu +\frac{2\pi\lambda}{e^2 r}\right)\,[dr^2+
r^2(d\theta^2 + \sin^2\! \theta \,d\phi^2) ]+
\left(\frac{2\pi\lambda}{e^2}\right)^2 
 \left(\mu+\frac{2\pi \lambda}{e^2 r}\right)^{-1}
  (d\psi+ \cos \theta\, d\phi)^2  \, .
\label{twomonometric}
\end{equation} 
Here $\mu$ is the reduced mass of the two fundamental monopoles, $r$,
$\theta$, and $\phi$ are the relative spatial coordinates, and $\psi$
is a relative $U(1)$ phase. 
The division by $Z$ indicates that there is an identification 
\begin{equation}
   (\chi,\psi)=(\chi+2\pi,\psi+
   \frac{4m_{\mbox{\boldmath $\gamma$}}}{m_{\mbox{\boldmath $\beta$}}+
   m_{\mbox{\boldmath $\gamma$}}}\pi)
\label{identification}
\end{equation} 
where $m_\bbeta$ and
$m_\bgamma$ are the masses of the two fundamental monopoles.  

    In fact, one can generalize from this case to that of maximal
symmetry breaking with an arbitrary number of distinct fundamental
monopoles.   In \cite{klee2} we
argued that for such moduli spaces the properties of the moduli space
metric for widely separated monopoles suggested that it was in fact
the exact metric over the whole moduli space.  Since then, Murray and
Chalmers \cite{murray} have offered proofs of this conjecture.

\section{Nonabelian unbroken symmetry: an SO(5) example}

As we have seen, generic values for the asymptotic Higgs field
give maximal symmetry breaking, with a nonabelian unbroken
symmetry emerging only for special values.  This suggests that one might
approach the latter case by studying the behavior of the MSB solutions
as the Higgs field approaches one of these special values.  However,
the $SU(3)$ example illustrated in Fig.~1 
shows that this limit is not always straightforward.  For example,
while the $\bbeta_1$ monopole of the MSB case seems to have a smooth
limit (the $\bbeta$ monopole) as the NUS case is approached, the same
is not true of the $\bbeta_2$ monopole, whose mass and core radius
tend to zero and infinity, respectively, in this limit.
Similarly, the transition from 
the eight-parameter family of MSB
solutions with magnetic charge proportional to $\bbeta_1^* +
\bbeta_2^*$ to the 
four-parameter
$\bbeta'$ monopole is certainly not smooth. 
In all of these cases the difficulties seem to be
associated with the large-distance behavior of the solutions, which is
complicated by the presence of massless nonabelian gauge fields.  This
suggests that the NUS limit might be better behaved when the
long-range magnetic fields are purely abelian; i.e., for solutions 
such that the ${\bf g} \cdot \bgamma_i$ all vanish.  In particular,
the MSB expressions for the mass, Eq.~(\ref{mass}), and for the
dimension of the moduli space, Eq.~(\ref{MSBmodes}), both remain valid
in the NUS limit in such cases.

The $SO(5)$ gauge theory provides a very nice example with which to
test this idea.  With the choice of Higgs field indicated in Fig.~2b,
the symmetry is broken to $SU(2)\times U(1)$.  According to
Eq.~(\ref{NUSmodes}), the minimal purely abelian magnetic charge,
given by $e{\bf g}/4\pi = \bbeta^* + \bgamma^*$, should have an
eight-dimensional moduli space of solutions.  It turns out that the
full eight-parameter family of solutions, all of which are spherically
symmetric, was explicitly found some time ago \cite{so5}.  Given these
solutions, it is a straightforward matter to find the moduli space
metric from Eq.~(\ref{modetometric}).  The corresponding solutions for
the maximally broken case (Fig.~2a) are composed of two fundamental
monopoles.  Although the explicit form of these solutions is not
known, we do know their moduli space and its metric; these were given
above in Eqs.~(\ref{space}-\ref{identification}).  Thus, we can check
by explicit calculation whether or not the NUS moduli space is indeed
the expected limit of that for the MSB case.

\vskip 5mm
\begin{center}
\leavevmode
\epsfysize =2in\epsfbox{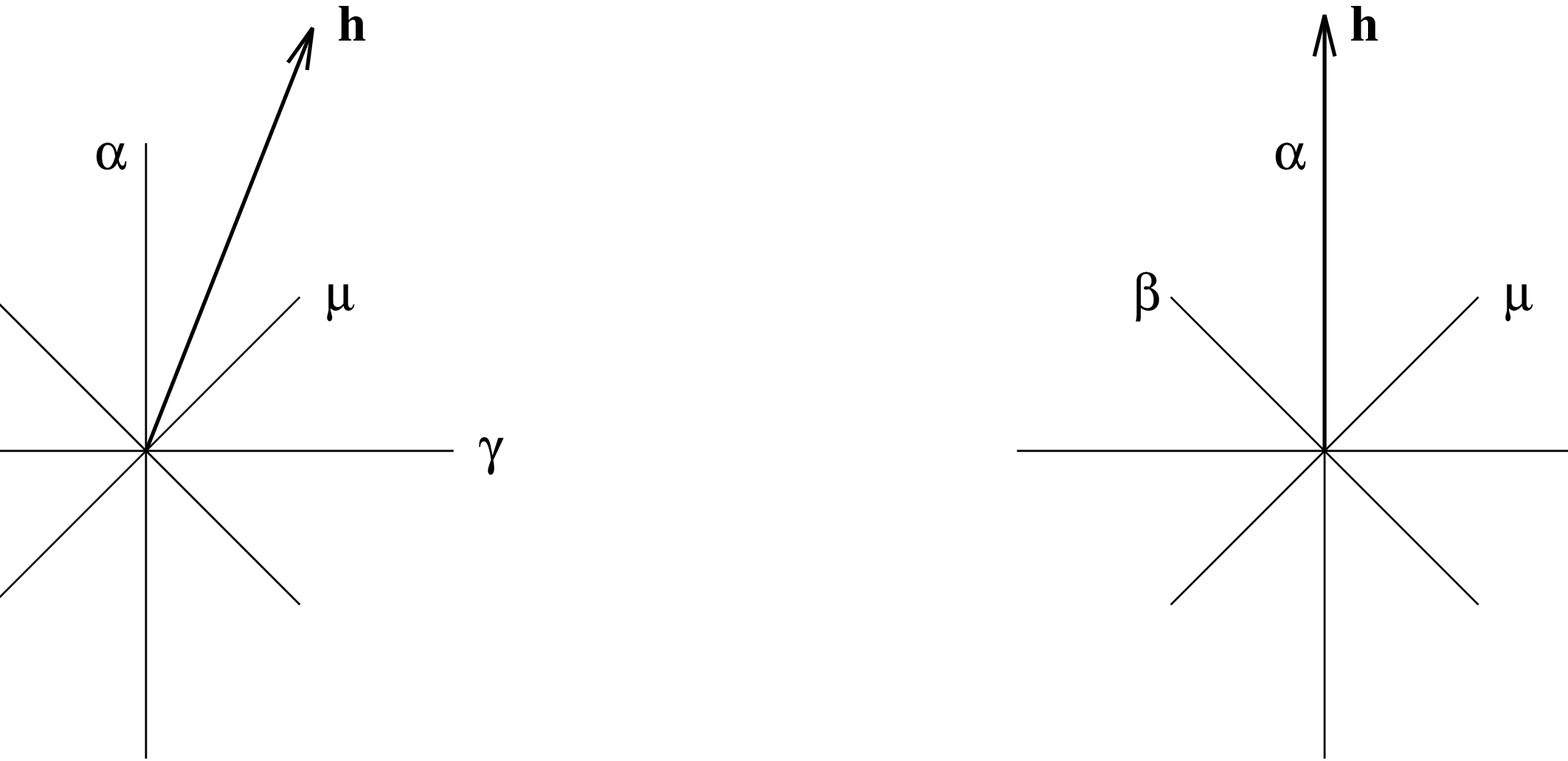}
\end{center}
\vskip 8mm
\begin{quote}
{\bf Figure 2:} {\small
The root diagram of $SO(5)$. With the Higgs vector $\bf h$ oriented as
in (a) the gauge symmetry is broken to $U(1)\times U(1)$, while with
the orientation in (b) the breaking is to $SU(2) \times U(1)$.}
\end{quote}
\vskip 0.38cm

Let us begin with the NUS solutions.  Three of the eight
parameters entering these specify the position of the center of mass,
while four others are obtained by applying global $SU(2)\times U(1)$
transformations to a solution.  The eighth parameter is not related to
a symmetry.  To see its significance we must examine the solutions in
detail.  First, we need some notation.  Any element of $P$ of the Lie
algebra of $SO(5)$ can be expressed in terms of two vectors ${\bf
P}_{(1)}$ and ${\bf P}_{(2)}$ and a $2\times 2$ matrix $P_{(3)}$ by
\begin{equation}
   P = {\bf P}_{(1)} \cdot {\bf t}({\mbox{\boldmath $\alpha$}}) 
       + {\bf P}_{(2)} \cdot {\bf t}({\mbox{\boldmath $\gamma$}}) 
       + {\rm tr}\, P_{(3)} M \, ,
\end{equation}
where ${\bf t}({\mbox{\boldmath $\alpha$}})$ and ${\bf t}
({\mbox{\boldmath $\gamma$}})$ are defined as in
Eq.~(\ref{tripletdef}) and 
\begin{equation}
  M = \frac{i }{ \sqrt{{\mbox{\boldmath $\beta$}}^2}} 
       \left(\matrix{E_{\mbox{\boldmath $\beta$}}  & 
      -E_{-{\mbox{\boldmath $\mu$}}} \cr
       E_{\mbox{\boldmath $\mu$}}  & \,\,\,
       E_{-{\mbox{\boldmath $\beta$}}}} \right)\, .
\end{equation}
The components of ${\bf P}_{(1)}$ are singlets under the
unbroken $SU(2)$, while ${\bf P}_{(2)}$ and $P_{(3)}$ transform as a
triplet and a pair of doublets.

With this notation, the family of solutions found in \cite{so5} can be
written as  
\begin{eqnarray}
   A^a_{i(1)} &=& \epsilon_{aim}{\hat r_m} A(r)  \qquad\qquad 
   \phi^a_{(1)} = {\hat r_a} H(r) \nonumber \\
   A^a_{i(2)} &=& \epsilon_{aim}{\hat r_m} G(r,a) \qquad\qquad
   \phi^a_{(2)} = {\hat r_a} G(r,a) \nonumber \\
   A_{i(3)} &=& \tau_i F(r,a) \qquad\qquad
   \phi_{(3)} = -i I F(r,a) \, .
\label{myansatz}
\end{eqnarray}
where $A(r)$ and $H(r)$ are the $SU(2)$ monopole functions given in
Eq.~(\ref{AHdef}) and
\begin{eqnarray}
   F(r,a) &=& { v \over \sqrt{8} \cosh (evr/2) }  L(r, a)^{1/2} \nonumber \\
   G(r,a) &=& A(r) L(r, a) \label{cloud}
\end{eqnarray}
with
\begin{equation}
   L(r, a) = \left[ 1 +  (r/a) \coth(evr/2) \right]^{-1} 
\end{equation}
and $v = {\bf h}\cdot \balpha$.

The parameter $a$ in these solutions has the dimensions of
length and can take on any positive real value.  It enters only in the
doublet and triplet components of fields.  Its effect on the triplet
components, proportional to $G(r,a)$, is particularly striking.  For $
1/ev \mathrel{\mathpalette\fun <} r \mathrel{\mathpalette\fun <} a$,
these fall as $1/r$, yielding the Coulomb magnetic field appropriate
to a nonabelian magnetic charge.  However, for larger distances the
vector potential falls as $1/r^2$, showing that the magnetic charge is
actually purely abelian.  One might describe 
these solutions as being
composed of a monopole core of radius $\sim 1/ev$ surrounded by a
``nonabelian cloud'' of radius $\sim a$.

Given these solutions, we can now
obtain the moduli space metric.  From symmetry considerations alone, we
see that the metric must be of the form
\begin{eqnarray}
    ds^2 &=& B(a) d{\bf x}^2 + C(a) d\chi^2 + I_1(a) da^2 
      + I_2(a) [d\theta^2 +\sin^2\! \theta \, d\phi^2 + (d\psi +\cos\theta
       \, d\phi)^2 ] \nonumber \\
      &\equiv & B(a) d{\bf x}^2 + C(a) d\chi^2 + I_1(a) da^2 + I_2(a)
        \, d\Omega_3^2  
\end{eqnarray} 
where $\bf x$ is the center of mass position and $\chi$ a $U(1)$ phase
angle, while the Euler angles $\theta$, $\phi$, and $\psi$ correspond
to the standard mapping of $SU(2)$ onto a three-sphere.  The BPS dyon
mass formula relates $B$ and $C$ to the monopole mass $M$, and shows
that they are both independent of $a$. 
The remaining metric functions, $I_1$ and $I_2$, can be
determined from the zero modes using Eq.~(\ref{modetometric}).  Making the
variation  $a \rightarrow a +\delta a$ in Eqs.~(\ref{myansatz}) yields
a zero mode that is already in background gauge and so can be directly
substituted into Eq.~(\ref{modetometric}) to give $g_{aa}= I_1=
4\pi/e^2\bgamma^2 a $.  To obtain the 
$SU(2)$ zero modes, we exploit the fact that the BPS
zero mode equations can be recast in the form of a Dirac equation
for $\psi(x) = I\delta \phi(x) + i \sigma_j \delta A_j(x)$ \cite{lowell}.
Multiplication of any solution $\psi$ on the right by a $2\times 2$
unitary matrix yields a new solution $\psi'$ that can be transformed
back to give a new BPS zero mode, already in background gauge.  In
particular, if $\psi_a$ is the Dirac solution obtained from the
$\delta a$ zero mode, then $\psi_a (i {\bf \hat n} \cdot \bsigma)$
yields a zero mode that corresponds to a gauge transformation with a 
gauge function of the form $f(r) \,{\bf \hat n}\cdot {\bf t}(\bgamma)$.
One finds 
that $f(\infty)=1/ea$, implying that the mode corresponding to
a shift $\delta a$ maps to an $SU(2)$ rotation by an angle $\delta a/a$,
which in turn implies that $I_2 = a^2 I_1$.  The net result is that
\begin{equation}
    ds^2 = M d{\bf x}^2 +  {16\pi^2 \over M e^4}d\chi^2 + 
     {4 \pi \over e^2 \bgamma^2} \left[{ da^2 \over a} 
      + a\, d\Omega_3^2 \right] .
\label{NUSmetric}
\end{equation} 
By making the change of variables $\rho =2\sqrt{a}$, this can be recast
in the form
\begin{equation}
    ds^2 = M d{\bf x}^2 +  {16\pi^2 \over M e^4}d\chi^2 + 
     {4 \pi \over e^2 \bgamma^2} \left[ d\rho^2  + {\rho^2\over 4}\,
   d\Omega_3^2 \right] \, .
\end{equation} 
The quantity in brackets is the metric for $R^4$ written in polar
coordinates (with the factor of $1/4$ arising from the
normalization of the Euler angles), and so the moduli space is the
flat manifold
\begin{equation}
    {\cal M} =  R^3 \times S^1 \times R^4 
\label{flatmanifold}
\end{equation} 
with the standard metric.  (The second factor is $S^1$ because of
the periodicity of $\chi$.)

The corresponding MSB moduli space was described in the
previous section.  The NUS case corresponds to the limit in which
$m_\bgamma$, and hence the reduced mass $\mu$, both vanish.  
In this limit the identification
(\ref{identification}) reduces to $(\chi,\psi) = (\chi +2\pi, \psi)$. The
division by $Z$ thus acts only on the $R^1$ factor, allowing us to
write 
\begin{equation}
   {\cal M} = R^3 \times S^1 \times {\cal M}_0  \, .
\end{equation}
Furthermore, if
we set
$\mu=0$ and use the fact that 
$\lambda = -2 \bbeta^* \cdot \bgamma^* = 2 /\bgamma^2$, the metric
(\ref{twomonometric}) for ${\cal M}_0$ becomes
\begin{equation}
   ds_0^2 =  \frac{4\pi}{e^2 \bgamma^2 r} \left( dr^2+ r^2\, d\Omega_3^2
\right) \, .
\end{equation} 
Comparing these results with Eqs.~(\ref{NUSmetric}) and
(\ref{flatmanifold}), we see that the NUS moduli space metric is
indeed the $m_\bgamma \rightarrow 0$ limit of that for the MSB case,
provided that 
intermonopole distance $r$ is identified with the cloud radius $a$.

Although the moduli space behaves smoothly as one case goes over into
the other, there is a curious change in the interpretation of its
coordinates.  The intermonopole distance becomes the radius of the
nonabelian cloud, while the angles specifying the relative spatial
orientation of the the two monopoles combine with their relative
$U(1)$ phase to give the global $SU(2)$ orientation of the solution.
Thus, the degrees of freedom of the $\bgamma$ monopole remain, but
they are no longer attributable to an isolated object.  Instead, they
describe the properties of a cloud that surrounds the $\bbeta$
monopole and cancels the nonabelian magnetic charge that the latter
acquires in the NUS limit.

We can see quite clearly how this comes about if we follow the
behavior of a generic MSB solution as the NUS limit is approached.
Thus, let us start with a two-monopole configuration in which the
intermonopole distance $r$ is much greater than the core radii
$R_\bbeta \sim (e^2 m_\bbeta)^{-1}$ and $R_\bgamma \sim (e^2
m_\bgamma)^{-1}$ of the two monopoles.  We can then take the NUS limit
by varying the asymptotic Higgs field so that $m_\bgamma \rightarrow
0$ while $m_\bbeta$ remains fixed.  If we had only a $\bgamma$
monopole, its core would expand without limit as we did this.
However, in the presence of the second monopole, the $\bgamma$ core
only expands until it reaches the $\bbeta$-monopole, at which point it
begins to evolve into a spherical cloud whose size is set by the
original intermonopole distance.  The reduced radius of the
$\bgamma$-monopole in the presence of a nearby $\bbeta$-monopole can
be understood in the same manner as that for the spherically symmetric
superposition of a $\bbeta_1$ and a $\bbeta_2$ $SU(3)$ monopole that
was noted previously.

\section{More complex examples}

In the $SO(5)$ example we saw how the two-monopole MSB solutions
evolved into spherically symmetric solutions with one massive and one
massless monopole in the limit of nonabelian symmetry breaking.
Because the exact NUS solutions were known, it was possible to verify
explicitly that the NUS moduli space was the expected limit of that
for the maximally broken case.  I will now discuss some more complex
examples.

The first of these, with two massive monopoles in the NUS limit,
arises in a theory with gauge group $SU(4)$.  The solutions are not
spherically symmetric, even in the NUS limit, and so it not surprising
that their explicit forms are not known.  However, because we have the
MSB moduli space metric, we check that it has the required isometries
in the NUS limit.

In this example the symmetry is broken to $U(1)\times SU(2) \times
U(1)$, with the $SU(2)$ factor corresponding to the middle root of the
Dynkin diagram in Fig.~3.  (This corresponds to a Higgs field
expectation value of the form $\Phi = {\rm diag}\,
(\phi_1,\phi_2,\phi_2,\phi_3)$ with $\phi_1 > \phi_2 > \phi_3$.)  If
\begin{equation}
    {\bf g} = {4\pi \over e} ( \bbeta_1^* + 
     \bgamma^* + \bbeta_2^*) \, , 
\end{equation}
then (1) ${\bf g} \cdot \bgamma=0$, implying that the asymptotic magnetic
field is purely abelian and that the moduli space should have a smooth
NUS limit, and (2) there is only one fundamental monopole of each
type, so the MSB moduli space is of the class for which the metric was
given in \cite{klee2}.  Examination of this metric shows that for the
generic MSB case it has an $SU(2)$ isometry corresponding to spatial
rotations and a $U(1)^3$ isometry corresponding to the unbroken gauge
symmetry.  However, the generator of one of these $U(1)$ factors
combines with two other vector fields, both of which become Killing
vectors in the NUS limit, to generate an $SU(2)$, so that in the 
NUS limit the $U(1)^3$ isometry is enlarged to the required $U(1)
\times SU(2) \times U(1)$.

\vskip 0mm
\begin{center}
\leavevmode
\epsfysize =0.6in\epsfbox{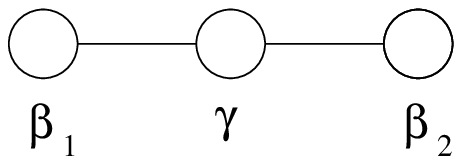}
\end{center}
\vskip 0mm
\begin{quote}
{\bf Figure 3:} {\small 
The Dynkin diagram of $SU(4)$, with the labelling of the simple roots
corresponding to symmetry breaking to $U(1)\times SU(2) \times U(1)$. }
\end{quote}
\vskip 0.38cm

The moduli space has twelve dimensions.  For the MSB case, a natural
choice of coordinates is given by three position variables and one
$U(1)$ phase for each of the three monopoles.  In the NUS limit, the
cores of the $\bbeta_1$ and $\bbeta_2$ monopoles remain finite and
(assuming that they are not too close) distinct.  Hence, the six
position variables specifying their locations should continue to be
good moduli space coordinates, as should their $U(1)$ phases.  The
overall $SU(2)$ orientation of the solution gives three more
coordinates.  This leaves just one other variable which, with the
$SO(5)$ example in mind, we might expect to in some way characterize a
nonabelian cloud surrounding the massive monopoles.

We saw in the $SO(5)$ example that the MSB intermonopole distance
became in the NUS limit a natural choice for an $SU(2)$-invariant
parameter describing the nonabelian cloud.  A nice generalization of
this occurs here.  Let $r_1$ and $r_2$ be the distances from the
$\bgamma$ monopole to the $\bbeta_1$ and $\bbeta_2$ monopoles,
respectively.  Their sum, $a=r_1+r_2$, (but not $r_1$ or $r_2$
separately) is left invariant by the vector fields that generate the
$SU(2)$ isometry and can be taken as the twelfth moduli space
coordinate.  Indeed, under the $SU(2)$ transformations generated by
these vector fields the orbit of the ``position'' of the $\bgamma$
monopole is the ellipsoid $r_1+r_2 =a$.

This example can be generalized to the case of $SU(N+2)$ broken to
$U(1)\times SU(N) \times U(1)$.  With the roots labelled as in Fig.~4,
the magnetic charge
\begin{equation}
\frac{e{\bf g}}{4\pi}={\mbox{\boldmath $\beta$}}_1^*+\sum_{j=2}^{N}
{\mbox{\boldmath $\gamma$}}_j^*+{\mbox{\boldmath $\beta$}}_{N+1}^*
\end{equation}
is orthogonal to the $\bgamma_j$'s that span the unbroken $SU(N)$.
Because there is only one fundamental monopole of each type, the MSB
metric of \cite{klee2} is again applicable.  Examining this metric,
one finds that there is a set of vector fields, generating a $U(1)
\times SU(N) \times U(1)$ algebra, that gives the required isometry in
the NUS limit.  The dimension of the moduli space is $4(N+1)$.  As
usual, the coordinates for maximal symmetry breaking can be taken to
be three position and one $U(1)$ variable for each of the $N+1$
fundamental monopoles.  With the nonabelian breaking, the positions
and $U(1)$ phases of the two monopoles that remain massive give eight
parameters.  The action of the unbroken $SU(N)$ gives additional
parameters; because a generic solution is left invariant by a $U(N-2)$
subgroup, there are only $[{\rm dim} \,\, SU(N) - {\rm dim}\, \,
U(N-2)] = 4N-5$ of these.  This leaves just one moduli space
coordinate to be specified; an $SU(N)$-invariant choice for this is $a
= \sum_1^N r_j$, where the $r_j$ are the distances between fundamental
monopoles whose corresponding simple roots are linked in the Dynkin
diagram of Fig.~4.

\vskip 5mm
\begin{center}
\leavevmode
\epsfysize =0.6in \epsfbox{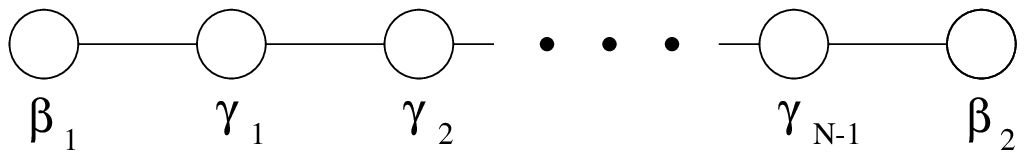}
\end{center}
\begin{quote}
{\bf Figure 4:} {\small 
The Dynkin diagram of $SU(N+2)$, with the labelling of the simple roots
corresponding to symmetry breaking to $U(1)\times SU(N) \times U(1)$. }
\end{quote}
\vskip 0.38cm

A purely abelian asymptotic field is also obtained if
\begin{equation}
\frac{e{\bf g}}{4\pi} =  N{\mbox{\boldmath $\beta$}}_1^*+ 
\sum_{j=1}^{j=N-1}(N-j){\mbox{\boldmath $\gamma$}}_{j+1}^*  \,.
\end{equation}
In the NUS limit this corresponds to a family of solutions containing
$N$ massive and $N(N-2)/2$ massless monopoles, with a moduli space of
dimension $2N(N+1)$.  The positions and $U(1)$ phases of the massive
monopoles give $4N$ of the coordinates, while the overall $SU(N)$
orientation gives $N^2-1$ more.  (The generic solution has no
invariance subgroup.)  This leaves $(N-1)^2$ parameters that describe
the gauge-invariant properties of the nonabelian cloud, which
evidently can have a much more complex structure than it did in the
previous examples.

The existence of two independent ``color-neutral'' magnetic charges
for this symmetry breaking (all others are linear combinations of
these two) is easily understood.  The massive monopoles
corresponding to $\bbeta_1$ and $\bbeta_2$ are objects that
transform under the fundamental representations $F$ and $\bar F$ of
the unbroken group, respectively.  The combinations given above
correspond to the fact that one can obtain an $SU(N)$ singlet either
with an $F$ and an $\bar F$, or from $N$ $F$'s.

\section{Concluding remarks}

In this talk I have shown how one can study monopoles with nonabelian
magnetic charges by following the behavior of purely abelian monopoles
as the asymptotic Higgs field is varied toward one of the special
values that leaves an nonabelian subgroup of the gauge symmetry
unbroken.  In the limit of nonabelian breaking, some of the abelian
monopoles remain massive, but acquire nonabelian components to their
magnetic charge.  Others, whose mass tends to zero in this limit,
evolve into a cloud that surrounds one or more massive monopoles and
cancels their nonabelian magnetic charge. Although they cease to exist
as distinct objects, their degrees of freedom survive as parameters
describing this cloud.

The analysis described here has been largely classical.  The effects
of quantum corrections remain to be investigated.  One would want to
see, for example, the effects of confinement on the nonabelian magnetic
charges.  It would also be desirable to go beyond the semiclassical
approximation and make contact with the results of Seiberg and Witten
\cite{seiberg}.

Perhaps most intriguing are the issues related to the duality
conjecture.  The monopoles that become massless in the limit of
nonabelian breaking are presumably the duals to the massless gauge
bosons.  A fuller understanding of these objects might well form the
basis for an approach to nonabelian interactions complementary to that
based on the perturbative gauge bosons.

\vskip .4cm

This work was supported in part by the U.S. Department of Energy.

\end{document}